\documentclass[%
 aps,prl,twocolumn,showpacs,superscriptaddress,
 groupedaddress,
nobibnotes,
 amsmath,amssymb,
floatfix,
]{revtex4-1}
\usepackage[mathscr]{eucal}

\usepackage{graphicx}
\usepackage{dcolumn}
\usepackage{bm}
\usepackage{epstopdf}
\usepackage{fourier} 
\usepackage{array}
\usepackage{makecell}
\usepackage{sidecap}
\usepackage[normalem]{ulem}
\usepackage[margin=.9in]{geometry}
\usepackage{amssymb, cancel, amsmath, amstext}
\usepackage[dvips]{epsfig}

\usepackage[ruled,vlined]{algorithm2e}

\SetCommentSty{mycommfont}

\usepackage{marginnote}
\usepackage{bbm}
\usepackage{caption}
\usepackage{subcaption}
\usepackage{colortbl}
\usepackage{mathrsfs}
\usepackage{bbm}
\usepackage{mathtools}

\usepackage{hyperref}

\usepackage{multirow}
\usepackage{sidecap}
\usepackage[normalem]{ulem}
\usepackage[margin=.9in]{geometry}
\usepackage{amssymb, cancel, amsmath, amstext}
\usepackage[all]{xy}
\usepackage{mathtools}

\usepackage{float}

\usepackage{color,soul}

\newcommand{\mylabel}[2]{#2\def\@currentlabel{#2}\label{#1}}

\usepackage{enumitem}
\newtheorem{Lemma}{Lemma}[section]

\newtheorem{Definition}[Lemma]{Definition}

\begin{document}

\preprint{}

\title{The Rising Sun Envelope Method:\\ an automatic and accurate peak location technique 
 for XANES measurements }

\author{Rafael Monteiro}
\email{monteirodasilva-rafael@aist.go.jp}
\affiliation{MathAM-OIL, AIST, c/o Advanced Institute for Materials Research, Tohoku University, Sendai, Japan}%

\author{Itsuki Miyazato$\*\*$}
\affiliation{Department of Chemistry, Hokkaido University, N-10 W-8, Sapporo 060-0810, Japan}
\affiliation{Center for Materials research by Information Integration (CMI$^2$),National Institute for Materials Science (NIMS), 1-2-1 Sengen, Tsukuba, Ibaraki 305-0047, Japan}%

\author{Keisuke Takahashi}

\email{$\*\*$miyazato@eng.hokudai.ac.jp, keisuke.takahashi@eng.hokudai.ac.jp}
\affiliation{Department of Chemistry, Hokkaido University, N-10 W-8, Sapporo 060-0810, Japan}
\affiliation{Center for Materials research by Information Integration (CMI$^2$),National Institute for Materials Science (NIMS), 1-2-1 Sengen, Tsukuba, Ibaraki 305-0047, Japan}%
\date{\today}

\begin{abstract}
The lack of theoretical understanding of X-Ray Absorption Near Edge Structure  (XANES) spectroscopy makes the development of analysis tools for its study a necessity. 
Here, an algorithm for judicious choice of local minima and maxima points of XANES spectrum (experimental or simulated) is proposed, without any loss of information on  peaks location nor on peak strength. We call it the  \textit{Rising Sun Envelope Method},  since it is based on successive regularizations of the spectral measurement that, according to parameter choices that are intrinsic to the measurements, keep peaks location and strength as invariants. This is the first method that finds peaks in XANES automatically, without depending on first derivative information. Nevertheless, a direct computation of Absorption-Edge is provided, where we avoid the issue inflection point computations based on the XANES second derivative, dealing instead with simpler computations of inflection points of higher quality cubic spline approximation. Besides applications of the algorithm to XANES, we illustrate  further applications in  Electron Energy Loss Spectroscopy (EELS) and Raman spectra. 

%
\end{abstract}

\maketitle

\section*{Introduction}
Atomic-Absorption Spectroscopy is a fundamental tool in Material Sciences for characterization of physical and chemical properties of materials \cite{koningsberger1988x}. 
In its foundations, the method is based on the photoelectric effect: upon interaction with  photons or other particles the material's atoms  absorb, emit, and reflect  incoming radiation to neighboring atoms, thus providing information about the material's atomic structure and its complexities \cite[\S 1-2]{bunkeR1010introduction}. 
In spite of its wide use, it is still a challenging task to extract meaningful and relevant information from spectra obtained in experimental measurements, and many theoretical questions in the field remain open.

One type of these measurements is known as X-Ray Absorption Near Edge Structure  (XANES) \cite{frenkel2012synchrotron,kuzmin2014exafs,newville2014fundamentals} observed in X-ray Absorption Fine Structure (XAFS),
where local structure and valence (or oxidation) state can be inferred from the way peaks amplify and shift horizontally when compared to peaks of a reference, non-oxidized, sample \cite{Zhao_structure,Belli,Peak_shift,lin2019highly,mao2017design}. Several approaches to valence determination and material oxidation exist: for instance,  Principal Components Analysis (PCA) is used to obtain \textit{valence state fingerprinting}, i.e., characterization of valence state by interpolating it with a mixture of pure-valence species \cite{PCA-valence}, a technique that is mathematically robust and powerful but rely on first derivative computations \cite{manceau2014estimating}, which lacks on mathematical rigor, for the spectra contains (non-differentiable) noise \cite{DEK}. Direct use of first derivatives is also common  \cite{First_derivative,ravel2005athena}. From a somewhat different perspective, Machine Learning (ML) techniques have been shown to be an effective method to unveil 3D structures in Nanoparticles \cite{ML-Timo-2019}, and characterize oxidation state by use of statistical methods \cite{zheng2018automated}; in the context of XANES, it has been applied with the aim of predicting oxidation state by learning the peak shift  \cite{miyazato2019automatic}. Nevertheless, all these techniques rely on peak location, derivative computations, or peak shift estimates, but up to now no method can locate a sequence of peaks in an accurate fashion. It is worth to point out that regularization through smoothing is successfully used in linear filtering and denoising of images, because it averages noise and keep low frequencies \cite{Mallat}; however, it smears peak location and their strength, which are the objects we are mostly interested at \cite[\S 6]{HaTiFr}.
      \newline
 \indent
 In this paper we take a step to address, and mitigate,  the aforementioned issues of characterization of peak location in XANES by means that are not rigorous (as differentiation methods) proposing a new method that does not smear out neither peak location nor peak strength. As it is the case with several tools of signal processing, the Rising Sun Envelope Method relies on resolution parameters that set up thresholds for peak location based on both their height and   how far apart they are.   One of the building blocks of the paper is called  the \textit{Rising Sun Envelope Method} \footnote{The idea is based on a construction by Riesz, in what became known as the Rising Sun Lemma \cite[Lemma 3.5]{Stein}, developed with the intent of understanding the pointwise behavior of continuous functions. In our case, we are interested in reducing the oscillation of the function in the sense of Def. \ref{def:osc}, which is exactly what the Rising Sun operator  does.}, in which an operator regularizes the XANES spectrum. The latter operator is applied successively in different energy range intervals, a  process  along which an invariant is kept at each step: in the corresponding interval, both the measurement and its regularization have their first peak located at the same point. In each step these intervals get smaller, yielding a  chained sequence of intervals and an increasing  sequence of peak locations in the energy range.
 
Towards the end of the paper several applications are given: approximated splines are constructed, and an automatic and a more accurate way to locate  inflection points is proposed. We remark that the latter do not rely on finite difference methods applied to the XANES measurement in order to find the second order derivatives, a computation that gives  highly oscillatory results due to noise but, in spite of its poor quality,  has been commonly used in the literature. Furthermore, the method provides a dimension reduction of the XANES measurement by judicious choice of interpolation points that can be then used for interpolation by polygonal, cubic splines , or clamped cubic splines glued up to first derivative. In the latter case, we assert the quality of our approximation by  comparing the inflection points obtained with the second order derivative  obtained by Athena, a well established software for XANES computations \cite{Athena}.  Finally, due to its generality, the Rising Sun envelope method can also be applied to other types of spectra,  a fact that we  illustrate with an application to Electron Energy Loss Spectroscopy (EELS) and Raman spectra.

 \section*{The Rising Sun and Valley of Shadows associated functions: an algorithm for peak location}\label{sec:rising_sun} Noise plagues spectral measurements with local maxima and minima, making the very distinction between  peaks and 
pure noise fluctuations in many ways relying on the experimentalist's subjective judgment. The  definitions in this section introduce two parameters that aim to both quantify and clarify the peak characterization. 

In what follows, we shall consider a spectral measurement $\mu(E)$, where $e_{-\infty}\leq E\leq e_{+\infty}$ denotes the energy range of measurement on a meshgrid $\delta n$, where $n$ is an integer and $\delta>0$ the meshgrid width. The continuous function $\mu(E)$
represents the material's absorption at energy level $E$. 
We shall denote the restriction of this mapping to a smaller energy rage $a\leq E \leq b$ by $\mu\Big|_{[a,b]}(E)$. Without any loss of generality, we can normalize the spectrum and consider $\displaystyle{\max_{e_{-\infty}\leq E\leq e_{+\infty}}\mu(E) - \min_{e_{-\infty}\leq E\leq e_{+\infty}}\mu(E)=1}$ and $\mu(e_{-\infty})=0$.

\begin{figure}[htb]
 \includegraphics[width=85mm]{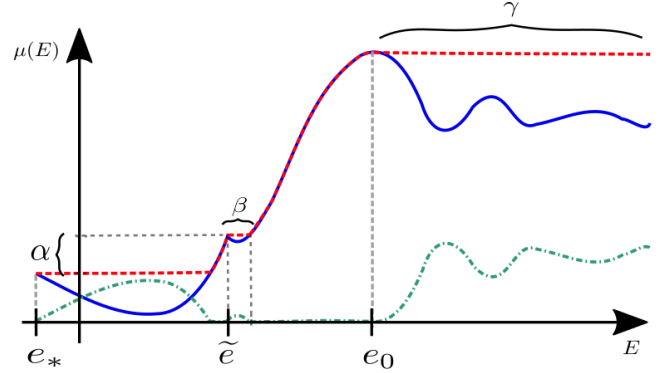}
 \caption{An illustration of the use of $(h_*, d_*)$ in peak finding:  $\mu(E)$ is represented as full line (blue). When looking for a local maximum with threshold $(h_*,d_*)$ relative to $(e_*,\mu(e_*))$, the first value in $\widetilde{e}$ is not considered a local maximum when $\alpha \leq h_*$; on the other hand, if $\alpha >h_*$, $\widetilde{e}$ is considered a local maximum only if $d_* <\beta$. The same analysis holds for $\overline{E}$, its relative height $\mu\left(\overline{E}\right) - \mu(e_*)$ and plateau size $\gamma$. The curve in  red represents the  Rising Sun function $\mathcal{R}_{\mu}(E)$ introduced in \eqref{rising_sun}, a valuable tool to verify the conditions in Def. \ref{def:rising_sun}.\label{fig:threshold_slecting_peaks}}
\end{figure}

\begin{Definition}[Peak criterion with thresholds $(h_*,d_*)$] \label{def:rising_sun} Given the a measurement $\mu(E)$ in energy range $e_{-\infty}\leq E\leq e_{+\infty}$ and fixed constants $h_*>0$ and $d_*>0$. Let $e_{-\infty}\leq e_*<\overline{E} \leq e_{+\infty}$. We say that $\overline{E}$ is a local maximum relative to $e_*$ with thresholds $(h_*,d_*)$ whenever
\begin{align}\label{def:loc_max1}
 \left\vert\mu\left(\overline{E}\right) - \mu(e_*)\right\vert >h_*, \quad\mbox{and} \quad \mu\left(\overline{E}\right) \geq \mu(E)
\end{align} 
holds for all  $e_{-\infty} \leq E\leq e_{+\infty}$ with $\vert E - \overline{E}\vert \leq d_*$ (see Figure \ref{fig:threshold_slecting_peaks}). Similarly, one says that point $e_*< \underline{E}\leq e_{+\infty}$ is a local minimum relative to $e_*$ with thresholds $(h_*,d_*)$ whenever
\begin{align}\label{def:loc_max1}
 \left\vert\mu(e_*) - \mu\left(\underline{E}\right) \right\vert>h_*, \quad \mbox{and} \quad \mu\left(\underline{E}\right) \leq \mu(E)
\end{align} 

holds for all  $e_{-\infty} \leq E\leq e_{+\infty}$ with $\vert E - \underline{E}\vert \leq d_*$.
\end{Definition}
The thresholds  $h_*>0$, $d_*>0$ are quantities used to discern local maxima/minima as \textit{pure noise} or \text{true peaks}. For instance,  taking the limit $h_* \to +\infty$ (resp., taking the limit $d_*\to  +\infty$), no local maxima/minima are found whereas, due to noise, in the limit $h_* \to 0$ (resp., $d_*\to \delta$) too many points are characterized as so. Thus, both thresholds must be tuned in order to capture intrinsic properties of the spectrum, like noise and relative distance between peaks; in that sense, both parameters play the role of \textit{resolutions}, as in some tools of signal processing \footnote{This is similar to the role of dilation and translation parameters in Wavelet theory with regards to resolution. The analogy is limited to the role of parameters, and do not extend beyond that because we are not doing time-frequency analysis \cite{Mallat,Lectures}. }. Furthermore, peak jumps display decaying properties, therefore it is important to take into account the relative dependence between successive maxima/minima, which is done by making each point a local maxima or minima with respect to the previous breakpoint. This remark leads to the following:

\begin{Definition}[M-breakpoints with $(h_*^{(n)},d_*^{(n)})$ threshold] \label{def:min_max_2}
Consider a  measurement $\mu(E)$ in the energy range $e_{-\infty}\leq E\leq e_{+\infty}$. Let $M >0$ be a fixed integer,  and $h_*^{(n)}>0$, $d_*^{(n)}>0$, where $0\leq n \leq M$. Consider the points 
 \begin{align}\label{chain}
  e_{-\infty} < e_0 <e_1 <... e_M \leq e_{+\infty}.
 \end{align}
We say that this is a  sequence \eqref{chain} of M-breakpoints with $(h_*^{(n)},d_*^{(n)})$ threshold whenever for all $n \geq1$, each $e_n$ with $n$ odd (resp., even) $e_n$ is a  local maximum (resp., local minimum) relative  to $e_{n-1}$ with threshold $(h_*^{(n)},d_*^{(n)})$ of the truncated spectrum $\mu\Big\vert_{[e_{n-1},e_{+\infty}]}(E)$, that is,  in the range $e_{n-1}\leq E\leq e_{+\infty}$.
\end{Definition}

\begin{figure}[htb]
 \includegraphics[width=85mm]{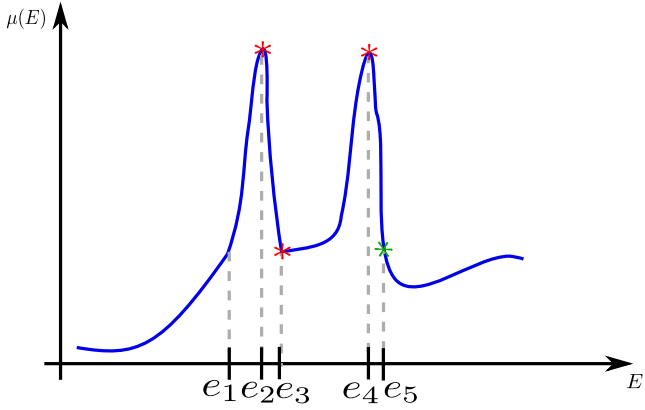}
 \caption{Sketch of XANES spectrum and breakpoints $e_1$, $e_3$ and $e_4$. Considering $e_3 - e_0 <d_* <e_4-e_1$, we have that $e_1$ and $e_3$  are local maxima with in a neighborhood of size $d_*$, while $e_3$  is not a local minimum in a neighborhood of same size due to the point $e_0$. Notice that the sharp decay of $\mu(E)$ in the range $e_1\leq E\leq e_3$ is controlled by the threshold $h_*$. Finally, note that if $e_5-e_1<d_*$ then the algorithm skip all the points in the range $e_1\leq E \leq e_4$ in the search for local minimum, being this one of the main reasons for readapting the threshold $d_*$ as the peak search happens; see also Figure \ref{fig:hidden_peak_trick} for more details. \label{fig:relative_local_max_min}}
\end{figure}
\begin{figure}[htb]
 \includegraphics[width=85mm]{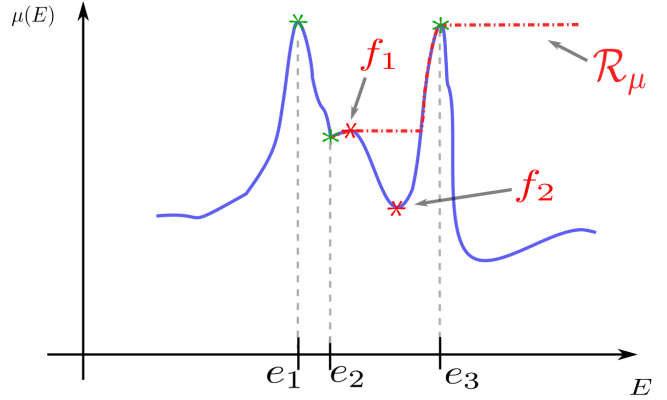}
 \caption{Sketch of the \textit{hidden peak trick} in a XANES spectrum $\mu(E)$. Considering an iteration from the local minimum $e_2$ with thresholds $(h^{(2)},d^{(2)})$, after which the algorithm constructs a Rising sun function $\mathcal{R}_{\mu}$. When $\vert\mu(f_1) - \mu(e_2)\vert<h^{(2)}<\vert\mu(e_3) - \mu(e_2)\vert$,  the algorithm sets $e_3$ as the next breakpoint, i.e., as local maximum, while skipping all the points in the range $e_2\leq E\leq e_3$. This is problematic, for it could clearly miss local minima as  $f_2$. The algorithm was implemented to check this case, allowing for two options: (i) $e_2$ is set as $f_2$, or (ii) $e_3$ is set as $f_1$; the algorithm runs from $e_3$ after this reassignment.  \label{fig:hidden_peak_trick}}
\end{figure}
\noindent
It is clear that a local maximum $e_0$ relative to $e_{-\infty}$ with threshold $(h_*^{(n)},d_*^{(n)})$ consists of a 1-breakpoint with $(h_*^{(n)},d_*^{(n)})$ threshold. The previous definition 
takes into account the order of the peaks and their sequential, pairwise, relative height. This step also implies that all the elements $e_n$ in \eqref{chain} with an  even index $n$ (resp., odd index $n$) are apart by a distance at least $d_*^{(n)}$. 

\begin{figure}[htb]
\includegraphics[width=85mm]{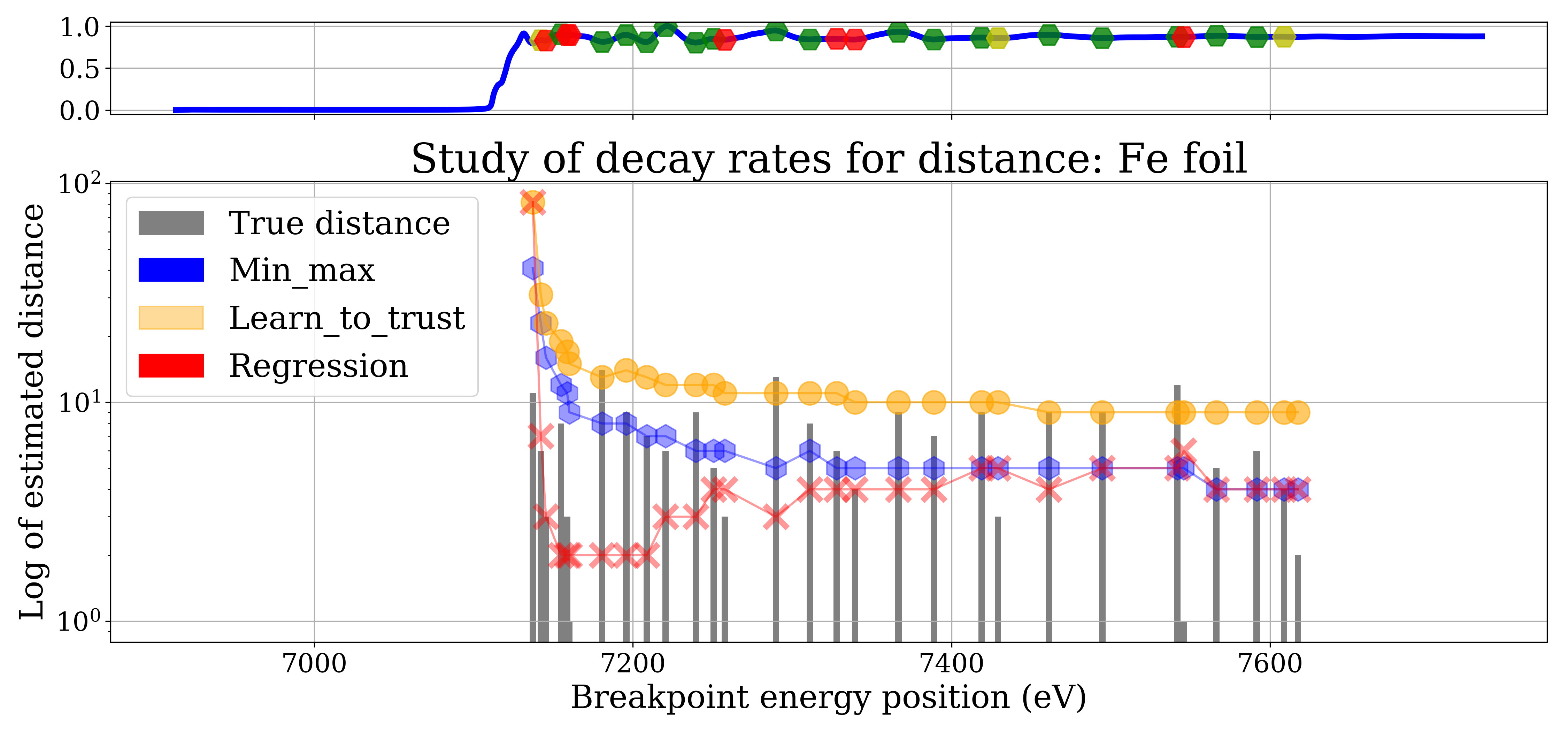}
 \caption{Some statistics for the threshold of distance  \textit{Fe foil}, which is then used as an estimate for the size of plateaus and, conveniently, as thresholds $d_*^{(n)}$. In all of which the estimate for $d_*^{(0)}$ is given by \eqref{estimate_d_0}. \label{fig:Statistics}}
\end{figure}

Several aspects of the previous definitions are worth of discussion. For simplicity, consider the thresholds  $(h_*^{(n)},d_*^{(n)})$ constants in $n$, assuming the value $(h_*,d_*)$:
\begin{enumerate}[label=\textnormal{(\arabic*)}]
\item[(C1)] \label{C1}  There is a trade-off between $h_*$ and $d_*$, and varying these quantities as the algorithm evolves is important to avoid local minima between two local maxima that would otherwise go undetected, as discussed in Figure \ref{fig:relative_local_max_min}. Three mechanisms are in place to avoid this problem: (a)  sharp variations in spectrum behavior are captured by the oscillation threshold (that is, $h_*$ large), (b) thresholds can vary throughout the search for breakpoints (see the  algorithm \ref{alg:_peak_location} and Figure \ref{fig:Statistics}), and the most effective of them all (c), called  the \textit{hidden peak tricks}, two techniques that introduce a spatial delay in the algorithm in order to verify whether its breakpoints are indeed local minimum/maximum, in some cases at risk of violating the peak threshold (see Figure \ref{fig:hidden_peak_trick}).

\item[(C2)] \label{C2} For any $n>1$, it is possible that the element $e_n$ in a M-breakpoint sequence with threshold $(h_*,d_*)$ is a local minimum/maximum  of $\mu\Big\vert_{[e_{n-1},e_{+\infty}]}(E)$, but not of $\mu\Big\vert_{[e_{-\infty},e_{+\infty}]}(E)$ (see Figure \ref{fig:relative_local_max_min}). However, for peaks found without using the \textit{hidden peak trick} it is true that the distance between any two of them that are successive local maxima or local minima 
is bigger than $d_*$; hence, they are all peaks with threshold  $(h_*,d_*)$ in $[e_0, e_{+\infty}]$. Last, one can deduce from Def. \ref{def:rising_sun} that the distance between peaks gives an estimate of the plateau sizes of the associated Rising Sun functions used to locate them;
\item[(C3)] \label{C3}The number $M$ is an upper bound on the number of peaks to be sought. In fact, the algorithm can stop before that many points are found: this is verified by checking if end of the range of measurement has been reached. Indeed, checking the Definition \ref{def:rising_sun} close to the endpoint can be an issue which we overcome by embedding the spectra in a higher dimensional space in a trivial fashion, that is, gluing the rightmost point to a constant in a continuous fashion; the latter does not affect peak location, peak height, nor the oscillation function (see Definition \ref{def:osc}). 

\end{enumerate}

\noindent
The following auxiliary constructions are useful to verify the properties in the previous definitions.
\begin{Definition}[Rising Sun and Valley of Shadows]
Given the XANES measurement $\mu(E)$ in the energy range $e_{-\infty}\leq E\leq e_{+\infty}$,   we define the Rising Sun operator
\begin{align}\label{rising_sun}
  \mathcal{R}_{\mu}(E) = \max_{e_{-\infty}\leq x \leq E} \mu(x),
\end{align}
where we call the function $\mathcal{R}_{\mu}(E)$ the Rising Sun function (associated to $\mu(E)$). We also define the Valley of Shadows operator,
\begin{align}\label{valley_of_shadows}
  \mathcal{V}_{\mu}(E) = \mathcal{R}_{\mu}(E) -  \mu(E),
\end{align}
where we call the mapping $\mathcal{V}_{\mu}(E)$ the Valley of Shadows function (associated to $\mu(E)$).
\end{Definition}
The following properties are an immediate consequence of the definition \ref{def:rising_sun} and the construction of the Rising Sun and Valley of Shadow functions $\mathcal{R}_{\mu}(E), \mathcal{V}_{\mu}(E)$, respectively.
\begin{enumerate}[label=\textnormal{(\arabic*)}]
 \item[(R1)] \label{R1} The function $\mathcal{R}_{\mu}(E)$ is non-decreasing. Furthermore, the inequality  $\mu(E) \leq \mathcal{R}_{\mu}(E)$ (equivalently,  $ 0 \leq \mathcal{V}_{S}[\mu](E)$) holds;

   \item[(R2)]\label{R2} Any non-decreasing function is invariant under the Rising Sun operator. In particular, the Rising Sun function is a fixed point of the Rising Sun operator, namely, $\mathcal{R}_{\mathcal{R}_{\mu}}(E) = \mathcal{R}_{\mu}(E)$;

   \item[(R3)]\label{R3} 
  Let the point $\overline{e}_1$ be the first local maximum of $\mu$ in the range $e_{-\infty}\leq E\leq e_{-\infty}$ with  threshold $(h_*,d_*)$. Then, $\overline{e}_1$ is the a local maximum of $\mathcal{R}_{\mu}$ in the range $e_{-\infty}\leq E\leq e_{-\infty}$ with  threshold $(h_*,d_*)$, that is, 
$$\mu(\overline{e}_1)=\mathcal{R}_{\mu}(\overline{e}_1) = \mathcal{R}_{\mu}(E), \quad  \overline{e}_1 \leq E < \overline{e}_1 + d_*.$$ In other words, the functions $\mu(E)$ and its associated Rising Sun function $\mathcal{R}_{\mu}$ have the same first maximum in the range $e_{-\infty}\leq E\leq e_{+\infty}$;

\item[(R4)]\label{R4}  Consider the first peak $\overline{e}_1$ with threshold $(h_*, d_*)$. Let $\overline{e}_2$ be the first local maximum with threshold $(h_*,d_*)$ of $\mathcal{V}_{\mu}(E)$ in the range $\overline{e}_1\leq E\leq e_{+\infty}$. Then
$(\overline{e}_2,\mu(\overline{e}_2))$ is a local minimum of $\mu\Big|_{[e_0,e_{+\infty}]}(E)$ with threshold $(h_*,d_*)$. Furthermore, 
$e_{-\infty}\leq e_0<e_1\leq e_{+\infty}$ is a sequence of  2-breakpoints with threshold $(h_*,d_*)$.
\end{enumerate}

\begin{figure}[htb]
 \includegraphics[width=85mm]{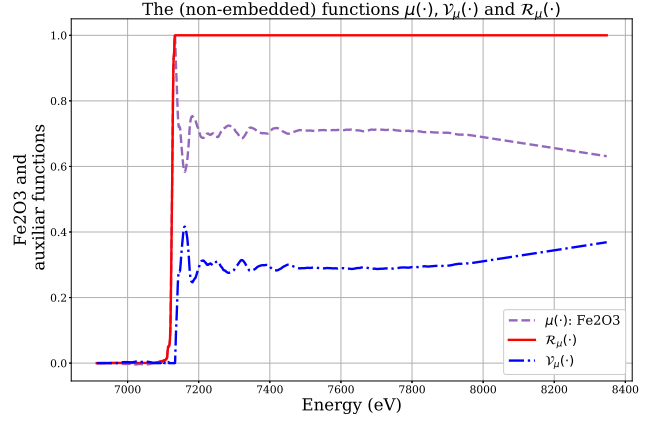}
 \caption{A spectral measurement $\mu(E)$ of $Fe_2O_3$,
  its associated Rising Sun  $\mathcal{R}_{\mu}(E)$ and Valley of Shadows functions $\mathcal{V}_{\mu}(E)$. Notice the correspondence between local mimina of $\mu(E)$ with local maxima of $\mathcal{V}_{\mu}(E)$.\label{risign_sun_valley_of_shadows}}
\end{figure}
\noindent
We carefully explain each of these  properties, as they are used in the implementation of Algorithm \ref{alg:_peak_location}: properties in (\hyperref[R1]{R1}) show that the Rising Sun function $\mathcal{R}_{\mu}(E)$ is the smallest non-decreasing majorant of $\mu(E)$; (\hyperref[R3]{R2}) shows that the family of non-decreasing functions is closed under the action of the Rising Sun operator; (\hyperref[R4]{R3}) asserts the invariance of the location and strength of the first local maximum with threshold $(h_*,d_*)$;  (\hyperref[R4]{R4}) provides the main idea of the algorithms, giving the foundation for the recursion in the sequence of smaller intervals $[e_{n-1},e_{+\infty}]\supset[e_{n},e_{+\infty}]$.
\paragraph*{Quantification of thresholds $(h_*,d_*)$.}\label{subsec:quantification} In the algorithm, the thresholds $(h_*^{(0)}, d_*^{(0)})$ are initialized before the peak search begins. Both quantities are highly dependent on the nature of the spectroscopy problem one deals with (see also Section \ref{subsec:eels_raman}): 
easier to quantify, $h_*$ denotes a the threshold for peak height aiming to distinguish peaks from random noise fluctuations; it is mostly dependent on the statistical error quantification or, more precisely, on the effect of noise on the regularity of $\mu(E)$. We estimate it using an oscillation function:

\begin{Definition}[Oscillation function]\label{def:osc}
 We define the oscillation threshold $h_*^{\mu}$ of   $\mu(E)$ as 
 \begin{align*}
h_*= \omega_*^{\mu}(\delta), \quad \text{for} \quad \omega_*^{\mu}(j \delta) = \max_{\vert E - E'\vert\leq j \delta } \vert \mu(E) - \mu(E')\vert.
 \end{align*}
 where $\omega_*^{\mu}(E)$ is called  \textit{oscillation function} of $\mu(E)$.
\end{Definition}
It is straightforward to show that $\omega_*^{\mathcal{R}_{\mu}}(j\delta)\leq \omega_*^{\mu}(j\delta)$, which amounts to saying that $\mathcal{R}_{\mu}(E)$ is more regular (less noisy) than the function $\mu(E)$ it is associated to; this can be seen as a \textit{denoising} effect, even though the analysis is not carried out in frequency space. 

In contrast to the previous case, the parameter  $d_*$ is heavily dependent on the distribution of peaks throughout the spectrum, which a priori is unknown.  By the construction of the Rising Sun function allied to  (\hyperref[C2]{C2}) we conclude that information about the distance between successive peaks provide estimates about the minimum size of the associated Rising Sun's plateaus. Thus, to initialize the code an estimate of this  distance is given by $d_*^{(0)}$, which we take to be
\begin{align}\label{estimate_d_0}
d_*^{(0)} = x_{1} - x_{0}, 
\end{align}
where we first compute $x_1$ as local maximum with threshold $(\lambda_1 h_*,\lambda_2 d_*)$ relative to $(e_{-\infty},\mu(e_{-\infty}))$; the quantities $\lambda_{1},\lambda_{2}$ are positive hyperparameters, which in computations for XANES were set as $\lambda_1=4$ and $\lambda_2=\frac{1}{4}$. It is clear from it's construction that we have $e_0 = x_1$.
Subsequently, we define $x_{0}$ as
\begin{align}
x_{0} = \max \left\{e_{-\infty}\leq E\leq e_{1} \,\Big|  \mu(E)  \leq \mu(-\infty) +h^{(0)}_* \right\}.
\end{align}
We can define what we call the \textit{min max} method:
\begin{equation}\label{min_max_method}
 \begin{split}
\mathrm{\alpha}_{\text{min max}}^{(n)}&=\frac{\max_{\tilde{x},\tilde{y} \in[e_{n},e_{+\infty}]}{\vert \mu(\tilde{x}) - \mu(\tilde{y}) \vert}}{\max_{x,y \in[e_{-\infty},e_{+\infty}]}{\vert \mu(x) - \mu(y) \vert}},\\ 
h_*^{(n+1)} &= \mathrm{\alpha}_{\text{min max}}^{(n)}\cdot h_*^{(n)},\\
\quad d_*^{(n+1)} &= \mathrm{\alpha}_{\text{min max}}^{(n)}\cdot \max\left\{\text{average}\left(d_*^{(k)}\right)_{0\leq k \leq n},2\right\}.
 \end{split}
\end{equation}
\noindent
Clearly, variations of the above give different results, specially for peaks that are harder to distinguish from pure noise spikes. An immediate objection to \eqref{min_max_method} is the  use of  peaks height's decay information  to infer the peak distance information, which give good results for small number of peaks, but is nonetheless far-fetched. For that reason other methods were also designed; for instance, the threshold formula is 
 \begin{align}\label{exponential_regression}
  d_*^{(n)} = \max\left\{\left(\lambda_3 e^{-n} + \frac{n}{2} \right)\frac{2L}{n+1}, 2\right\},
 \end{align}
 where $\lambda_3$ is a hyperparameter. Each method estimates the quantity $L$ in a different way:

\begin{enumerate}
 \item[i.] (\textit{Regression}) Given the location of i peaks, estimate the plateau size of the next one by doing the a regression of type 
 \begin{align}\label{L_regression}
  L = \alpha_d \exp\left\{\beta_d (i+1)\right\},
 \end{align}
where $\alpha_d$ and $\beta_d$ are estimated using the points  $\left(j, d_*^{(j)}\right)$ for $0\leq j \leq i$. 

\item[ii.] (\textit{Learn to trust}) Given the location of i peaks, estimate the the quantity L as 
\begin{align}\label{L_learn_to_trust}
L = \text{average}\left(d_*^{(k)}\right)_{0\leq k \leq n}. 
\end{align}
\end{enumerate}
Assuming $L$ fixed, note that when $n=0$ the estimate \eqref{exponential_regression} gives 
$$d_*^{(0)} = \max\left\{\lambda_3L,2 \right\};$$
on the other hand, as $n\to +\infty$
$$\lim_{n \to +\infty}d_*^{(n)} =  \max\left\{L,2 \right\}.$$
In this case, the initial guess  given by the hyperparameter $\lambda_3$ plays an important role in the search of peaks immediately after the first located peak; however, this guess lose importance as the method gather more data from previous peaks,  eventually \textit{learning form it}.  In Fig. \ref{fig:Statistics} we plot a comparison between several methods to estimate the distance; see \cite{github} for further details. 

\begin{figure}[htb]
\includegraphics[width=85mm]{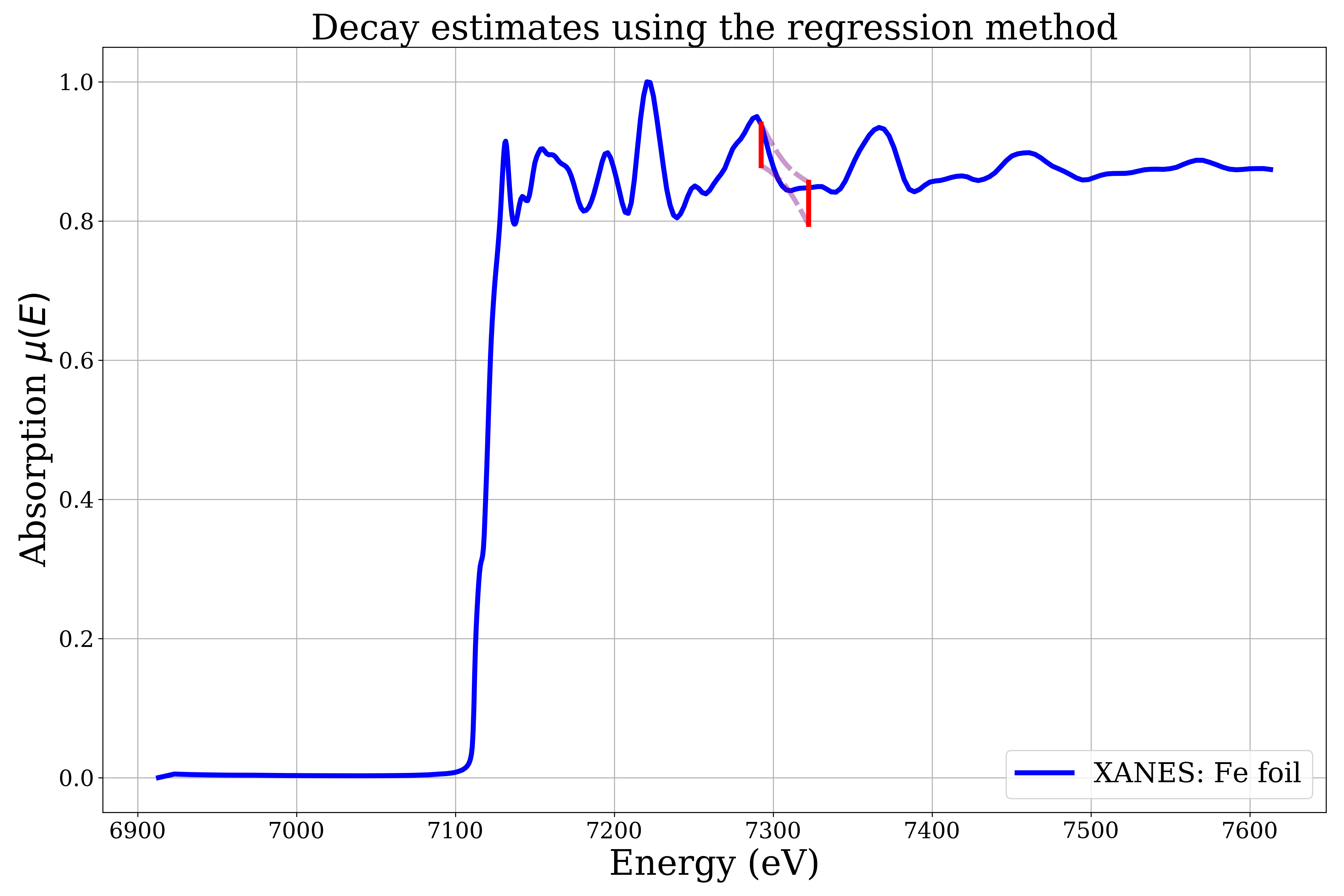}
 \caption{An illustration of the \textit{regression method} applied to Fe foil. The error curves  \eqref{regression:error_curves} are in purple, while the red curves give the error range at the beginning and at the end of the regression range.\label{fig:regression_method}}
\end{figure}

\noindent
A few more words about the \textit{regression method} are necessary: in the way it has been applied to estimate $d_*^{(n+1)}$ from $\left(h_*^{(j)},d_*^{(j)}\right)_{1\leq j \leq n}$ it can also be applied to estimate $h_*^{(n+1)}$. Indeed, at the $n$th peak we would like to find a good candidate for the threshold $h_*^{(n+1)}$; recall that the location of the $(n+1)$-th peak is unknown. We proceed as follows: a linear regression of $\mu\Big|_{[e_{n}, e_n + 2d_*^{(n+1)}]}$  on $E\Big|_{[e_{n}, e_n + 2 d_*^{(n+1)}]}$, where $d_*^{(n+1)}$  is computed using \eqref{exponential_regression} with $L$ as in \eqref{L_regression}. We estimate and write the error in the regression  as $\widehat{\sigma}(E)$ (see \cite[\S 8, Eq. (8.4)]{HaTiFr}), from which we obtain two curves,
\begin{equation}\label{regression:error_curves}
 \widehat{\mu^{(\pm)}}(E) =\underbrace{ \beta_0 + \beta_1E}_{\text{linear prediction}} \pm  \lambda_4\cdot\widehat{\sigma}(E),
\end{equation}
for $e_{n} \leq E \leq e_n + 2d_*^{(n+1)}$ and $\lambda_4$ a hyperparameter (in our case set  as $\lambda_4 =3$ throughout computations). Now, define $\displaystyle{S_{\text{regression}} \equiv \max_{\tilde{x},\tilde{y} \in[e_{n},e_{n} +2d_*^{(n+1)}]}\vert \widehat{\mu^{(+)}}(\tilde{x}) - \widehat{\mu^{(-)}}(\tilde{y}) \vert }$, which gives an estimate on the jump in intensity between the $n$-th peak (already known) and the $(n+1)$-th peak (unknown). We have then the \textit{regression method}, a slight variation of \eqref{min_max_method},
\begin{equation}\label{regression_method}
 \begin{split}
\mathrm{\alpha}_{\text{regression}}^{(n)}&=\frac{S_{\text{regression}}}{\text{average of previous jumps}},\\ 
h_*^{(n+1)} &= \mathrm{\alpha}_{\text{regression}}^{(n)}\cdot h_*^{(0)}.
 \end{split}
\end{equation}
The method is illustrated in Fig. \ref{fig:regression_method}.

\paragraph*{Searching on the other side of the Near-Edge peak: the reflection method.}\label{subsec:reflection}

\begin{figure}[htb]
 \includegraphics[width=85mm]{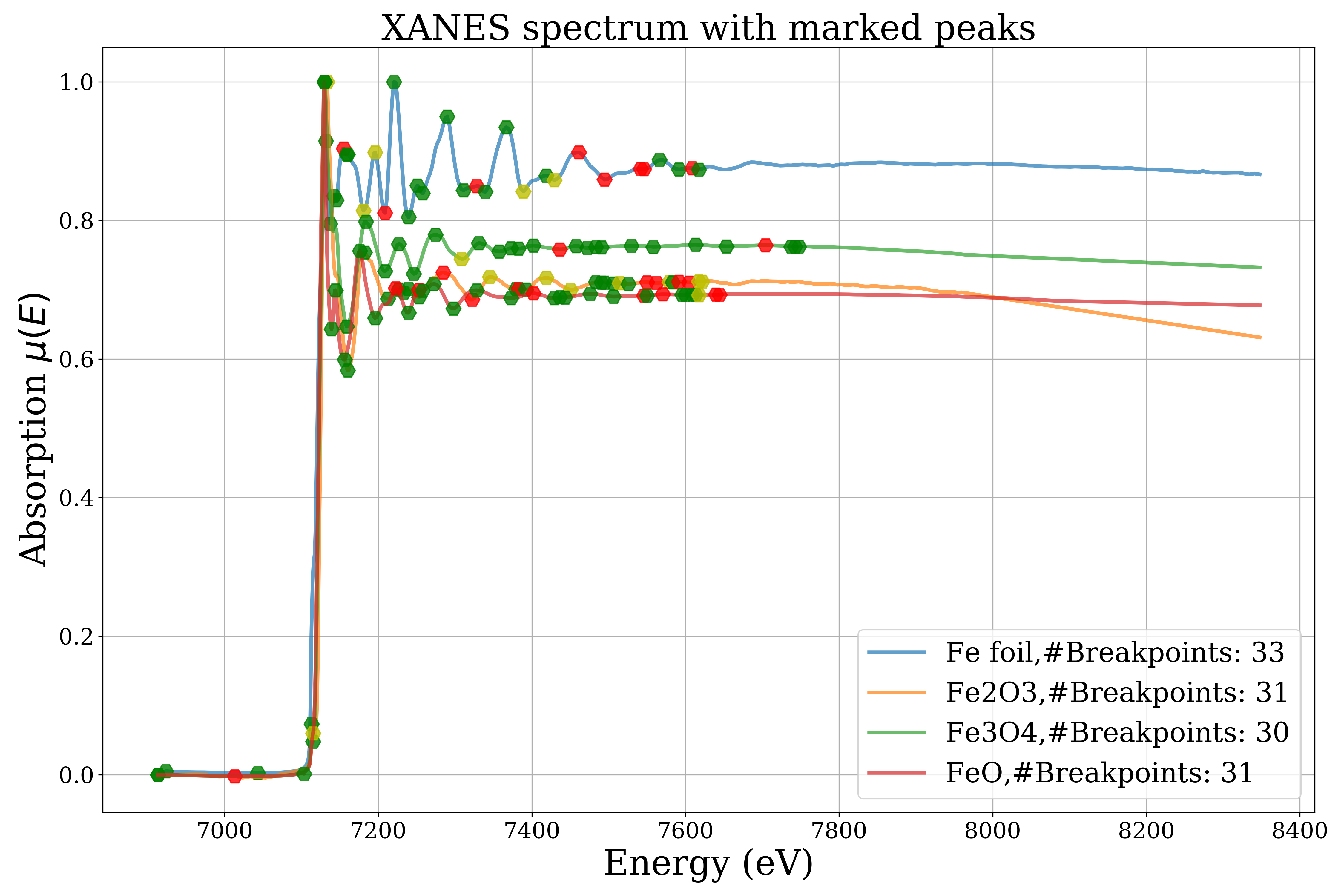}
 \caption{An example of peak localization in Fe samples. For peaks on the right (resp. left) of the 0th peak, i.e., peak $e_n$ with  $n>0$ (resp. $n<0$), the  color of the marker indicates the ratio between oscillation in $\mu\Big|_{[e_{n},e_{+\infty}]}$ (resp. $\mu\Big|_{[e_{-\infty}, e_{n}]}$ and the amount of jump to the previous peak; similarly, the color on the 0th peak denotes  the ratio between oscillation in the whole energy range $[e_{-\infty}, e_{+\infty}]$ and its jump (or estimated jump) to previous peak: the color is green whenever this ratio is less than 1, yellow for ratio in between 1 and 2, and red whenever the ratio is above 3. This ratio can be used as a measure of "peak relevance", allowing one to introduce scores to classify the quality of the found peaks. \label{fig:other_side_with_marked_peaks}}
\end{figure}

By construction,  the algorithm starts at  $e_0$,  somehow ignoring any peak to the left of it. This is somewhat intentional, and is done for two reasons: first, away from the peak-edge (the ``0th peak"), peaks on both sides decay at different rates; second, the algorithm can use itself in a recursive fashion to search for peaks on the left side once it has finished the search on the right side. The latter is carried out by reflection of the spectrum  about the peak $e_0$,   $\widetilde{\mu}(E) \equiv \mu( e_{0}+e_{-\infty} -E)$ and restricting the range of search to $e_{-\infty}\leq E\leq e_{-\infty}+e_0$; the initial thresholds $\widetilde{h}_*^{(0)}$ and  $\widetilde{d}_*^{(0)}$ are estimated from $h_*^{(n)}$ and  $d_*^{(n)}$. 
The output is shown in Figure \ref{fig:other_side_with_marked_peaks}.

\paragraph*{Pathological measurements, threshold readjustment,  and the role of noise.}\label{subsec:pathological}

We remark that the algorithm takes advantage of the presence of noise:  in the case of an unsuccessful search for peaks with thresholds $(h_*,d_*)$, the algorithm readjust the quantities to 
$$ h_*\leftarrow  \lambda_5 h_*, \quad \text{and} \quad d_* \leftarrow d_* -1,$$
either in an alternate fashion or simultaneously (also a hyperparameter, we set $\lambda_5 = 0.9$ throughout our computations); see \cite{github} for further details.
In constrast, the algorithm would stop in the first peak of a polygonal curve that has plateaus bigger than $d_*$; the implementation takes these as pathological cases, for they are not expected in XANES. Moreover,

\section*{Further applications}

\paragraph{A dimension reduction method using judiciously chosen  breakpoints for interpolation.}\label{subsec:dimension_reduc}

\begin{figure}[htb]
 \includegraphics[width=85mm]{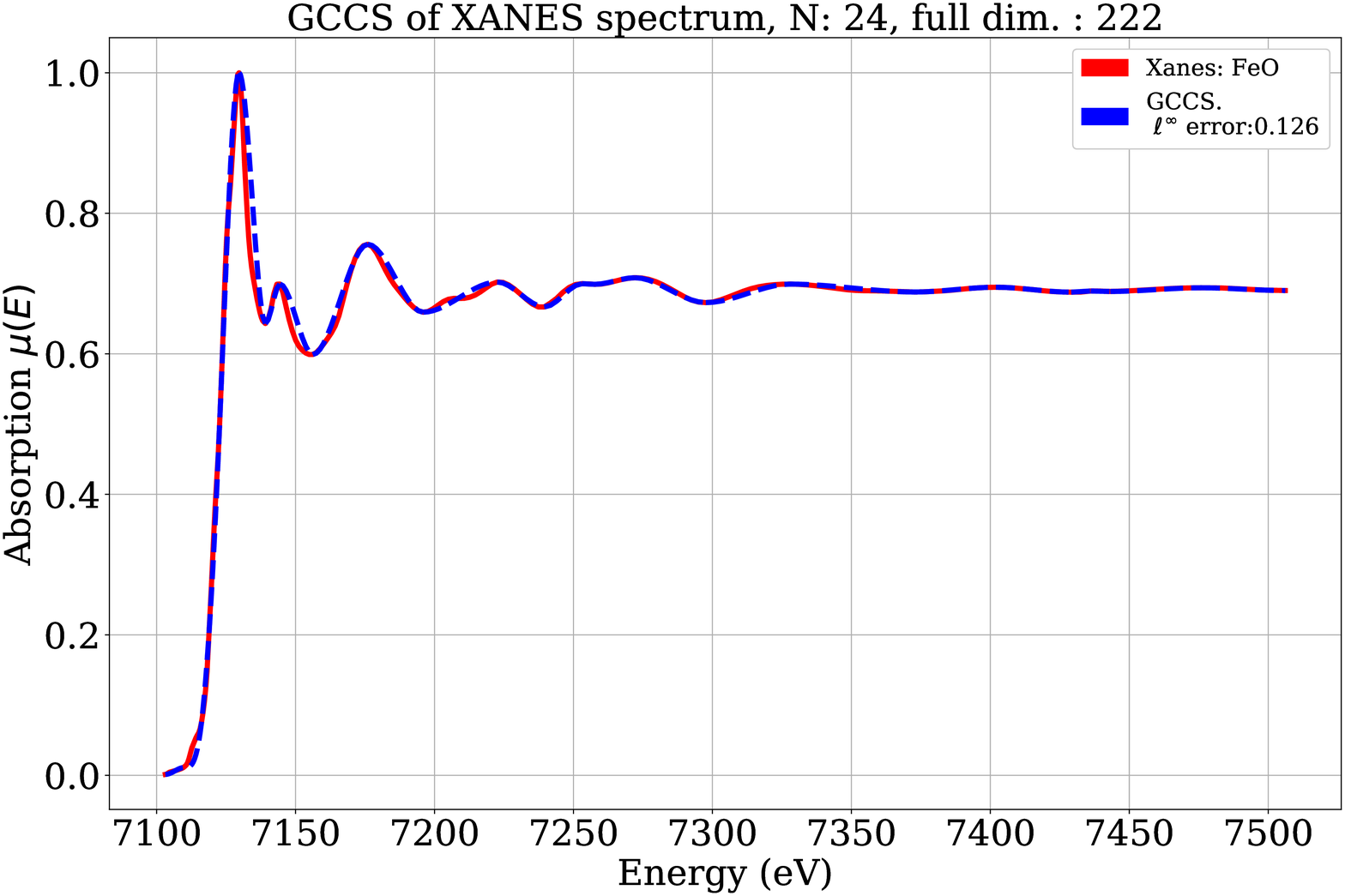}
 \caption{Glued Clamped Cubic Spline for $FeO$, with 24 interpolation points, using judiciously chosen interpolating points chosen by the  Rising Sun Envelope Method. \label{fig:clamped_FeO}}
\end{figure}

\begin{figure}[htb]
 \includegraphics[width=85mm]{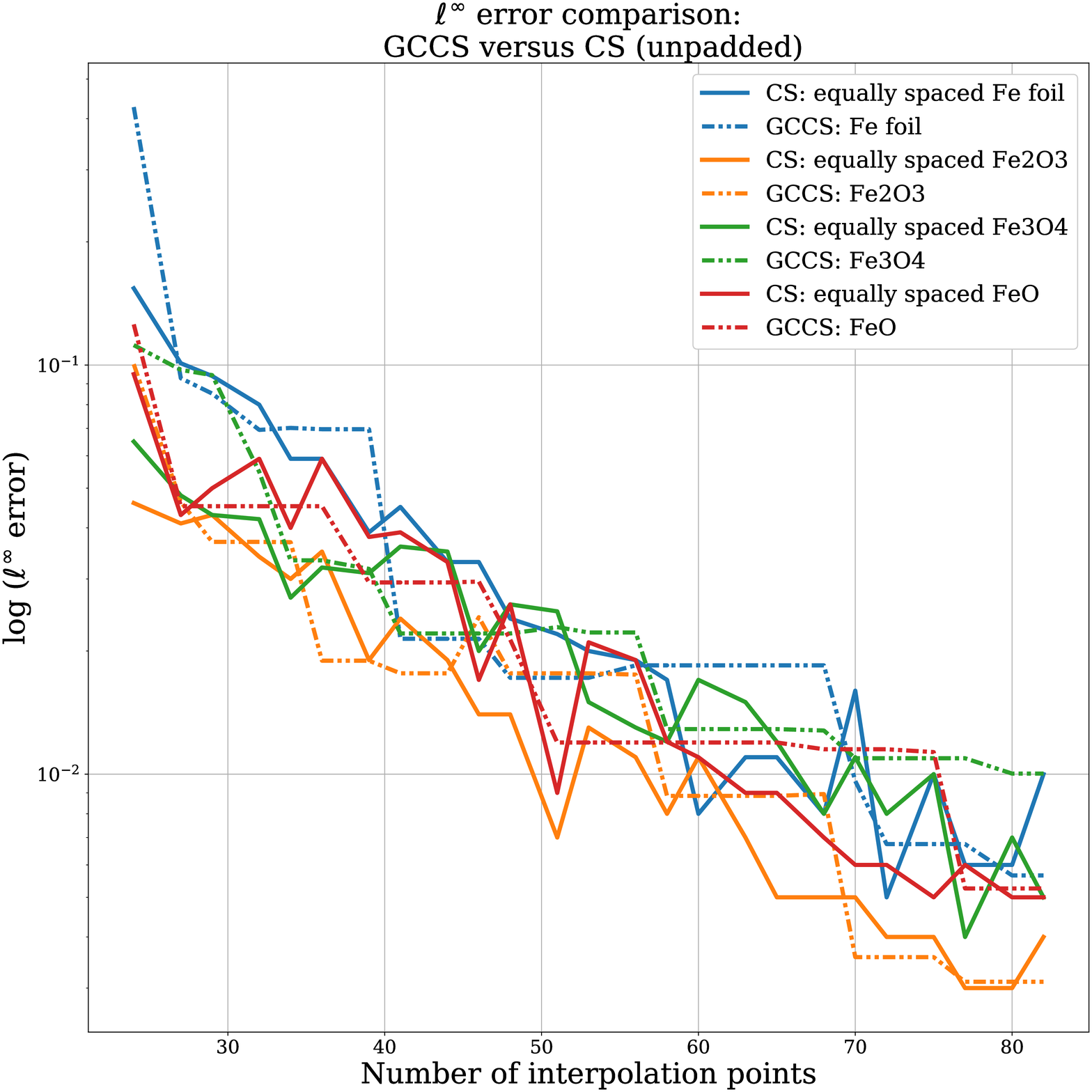}
 \caption{The judicious choice of interpolation points provided by our algorithm leads to better approximation result in $\ell^{\infty}$ norm, and also to exact location of peaks. Comparison between Glued Clamped Cubic Spline (GCCS) method and Cubic Splines (CS).\label{fig:error}}
\end{figure}

 With the breakpoints $e_{-\infty}\leq e_{-M_{-\infty}}\leq \cdots\leq  e_0 \leq e_1 \leq \cdots \leq e_{M_{\infty}}\leq e_{+\infty}$ in hands we now look for a lower dimensional representation of $\mu(E)$ in the energy range $e_{-\infty}\leq E \leq e_{-\infty}$. That is, given the points  $\displaystyle{\left(e_{m}, \mu(e_{m})\right)_{-M_{-\infty}\leq m \leq M_{\infty}}}$ we can choose several methods to interpolate the given spectrum $\mu(E)$ (see Figure \ref{fig:clamped_FeO}). The number of points in the interpolation can be increased by further  refinement of the intervals $\left[e_m,e_{m+1}\right]$, for $m \in \{-M_{-\infty},\ldots ,M_{\infty}-1\}$. 
The breakpoints $e_{m}$, for $-M_{-\infty}\leq m \leq M_{\infty}$ provide a partition of the energy range $e_{-\infty}\leq E\leq e_{+\infty}$ into intervals of \textit{almost monotonic behavior}. In passing, it allows for the use of interpolation methods that exploit this fact. Indeed, in between two successive breakpoints we interpolated a cubic clamped spline, that is, a spline in  with zero derivative at its endpoints; we call it Glued Clamped Cubic Spline (GCCS) the resulting concatenation of the these clamped cubic splines along the interpolated intervals, a result that  provides a better quality of approximation when compared to other types of interpolations (see Figure \ref{fig:error}). Other techniques have also been implemented; for instance: padding the endpoints of the XANES spectrum with constants was used to tame the wild behavior of interpolation near the boundaries \cite[\S 5.2.1]{HaTiFr}, where the spectrum was padded, interpolated and then truncated to the relevant energy region. 

\begin{figure}[htb]
 \includegraphics[width=85mm]{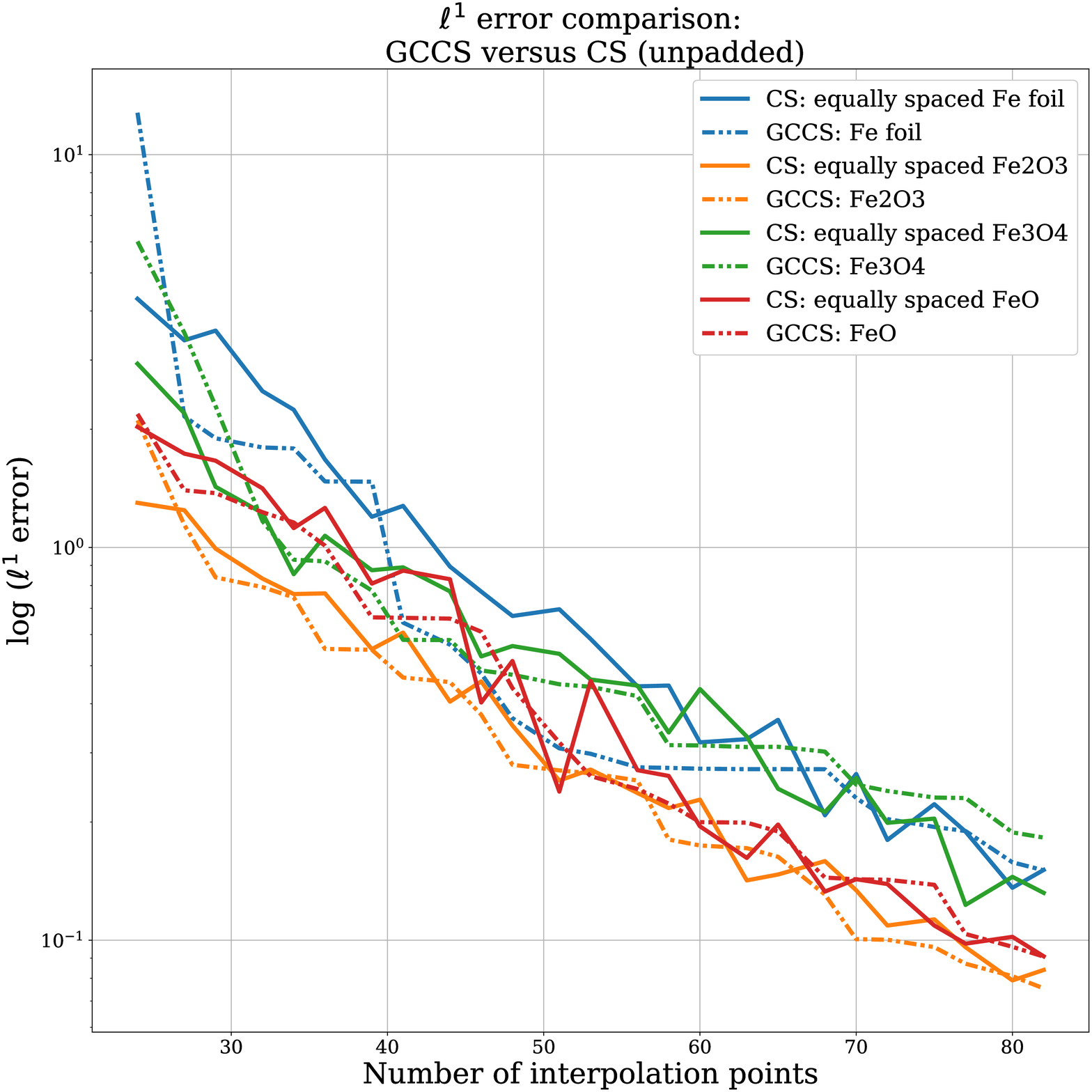}
 \caption{Study of approximation error of interpolation in   $\ell^{1}$ norm comparing the Glued Clamped Cubic Splines (GCCS) method to Cubic Splines (CS) interpolation using equally spaced meshgrid.  \label{fig:error_l_1}}
\end{figure}

In Figures \ref{fig:error} (resp.,  \ref{fig:error_l_1}) we show the $\ell^{\infty}$  interpolation error, i,.e., maximum of absolute difference, (resp., $\ell^{1}$ interpolation error, i.e., sum of the absolute differences). WE compare our result to other Cubic Spline interpolations with equally spaced meshgrids with same number or points. The error  suing Glued Clamper Cubic Splines (GCCS) is mostly smaller in lower dimensions, while it gets similar as the dimension (number of interpolation points) increases.

\begin{figure}[htb]
 \includegraphics[width=85mm]{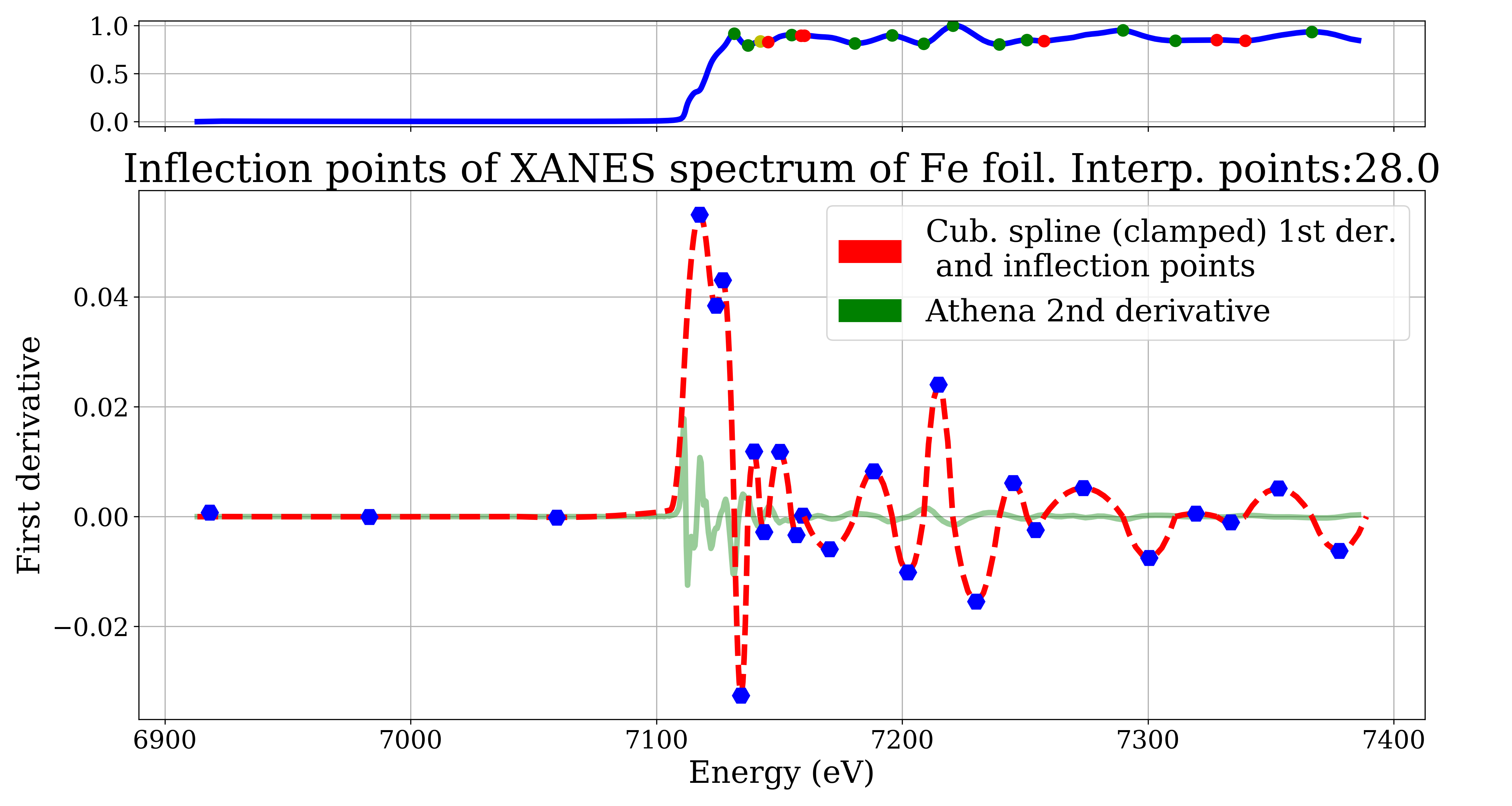}
 \caption{Inflection points of XANES measurement, computed from approximation curved. In comparison, the second derivative of the XANES spectrum found using \textit{Athena}  \cite{Athena} is also given, noticeably more susceptible to noise.\label{fig:First_derivative}}
\end{figure}

\paragraph{Inflection point calculation.}\label{subsec:inflect point}
In figure \ref{fig:First_derivative} the first derivative of the interpolant approximation. For peak location purposes, plotting the first derivative is unnecessary because the Rising Sun Envelope Method, for given thresholds, finds exact peak location. Nevertheless, it is common in the literature to associate the highest value of the first derivative to the point of highest absorption, which is also called  the  \textit{Absorption Edge}.  As an  inflection point of the XANES measurement, the second derivative of the measurement is commonly  computed in spite of its poor quality due to noise (see Figure \ref{fig:First_derivative}). For comparison, we plot the first derivative of our interpolant and all its inflection points, which are robust and more amenable to computation due to the good differentiability properties of the clamped splines in each interval they are defined in. The poor information given by the second derivative of the spectrum prevents much information to be obtained from it whereas,  in contrast, the quality of the inflection points found using the clamped splines approximation. We highlight that the approximation used in Figure \ref{fig:First_derivative} requires only 28 interpolation points out of $418$, that is, only approximately  $6\%$  of the whole  spectrum information was used.

\paragraph{Applications to EELS and Raman spectra.}\label{subsec:eels_raman}

\begin{figure}[htb]
 \includegraphics[width=85mm]{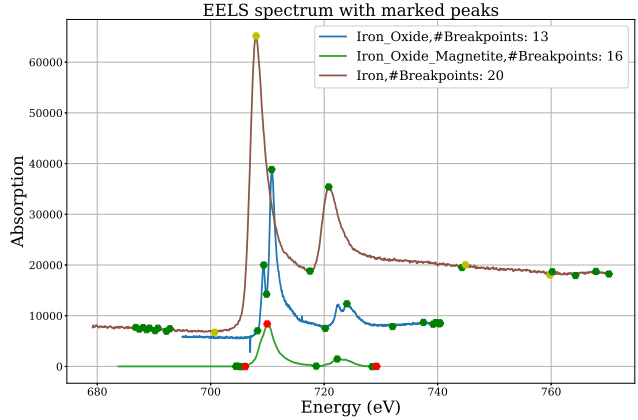}
 \caption{EELS spectra or Iron, Iron oxide, and  Iron oxide magnetite with marked peaks; data source \cite{EELS}.\label{fig:EELS}}
\end{figure}

\begin{figure}[htb]
 \includegraphics[width=85mm]{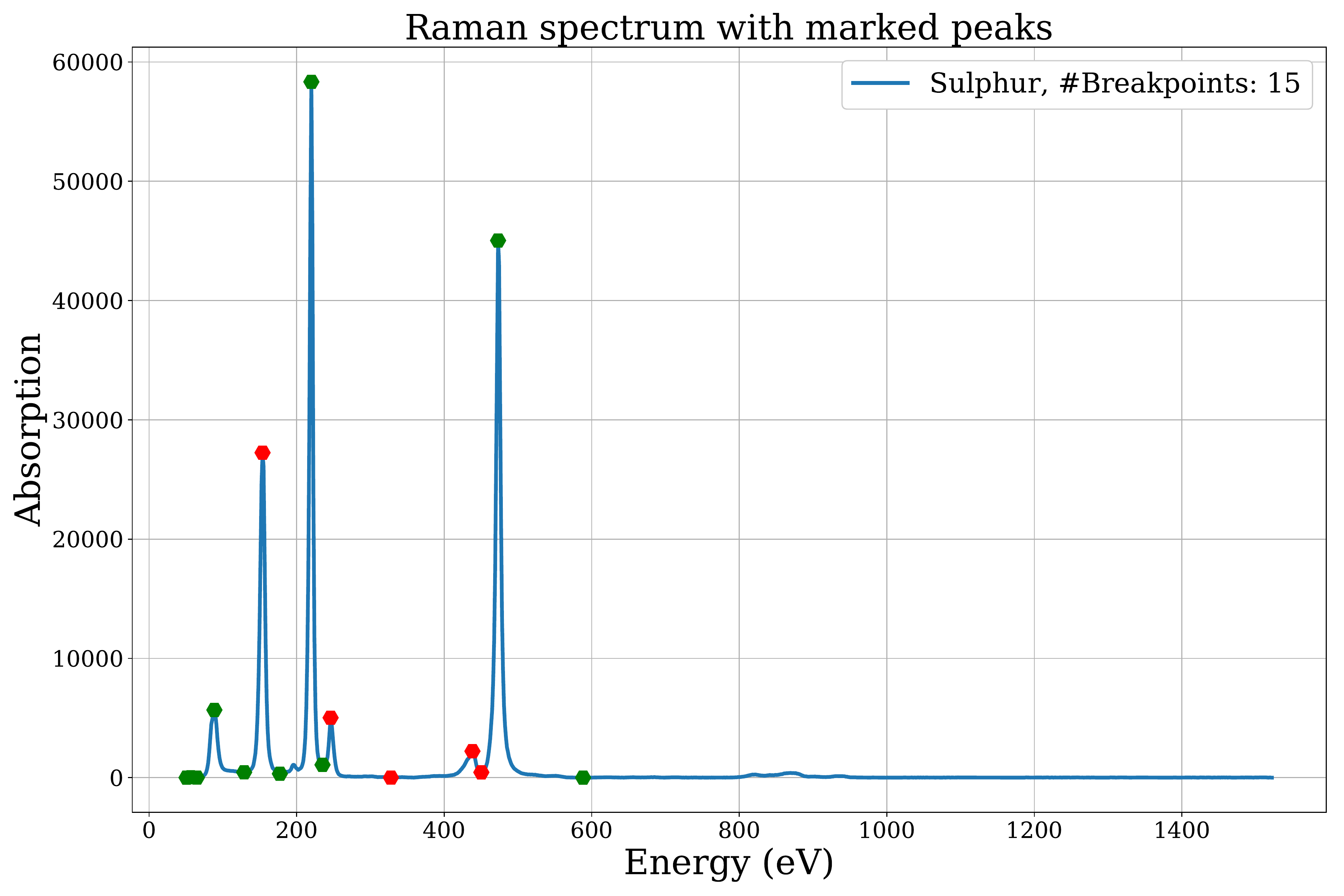}
 \caption{Raman spectra of depolarized Sulphur with marked peaks; data source\cite{RUFF}.  \label{fig:Raman}}
\end{figure}

The characterization of peaks shift and its intensity in Electron Energy Loss Spectroscopy (EELS) have a similar reason as that in XANES, therefore it is a natural object for validation of our code (see Figure \ref{fig:EELS}).
In Fig. \ref{fig:Raman} the Rising Sun Envelope Method is applied to Raman Spectra of Sulphur.  In Raman Spectroscopy peak location is used as a initial that precedes  Gaussian, or other type of, fitting  \cite{larkin2017infrared}; in this case, knowledge of peak position and its intensity helps in qualitative and quantitative comparison between spectrum and data libraries, which is a fundamental step in Raman spectroscopy.  It is worth to point out  that in Raman the distribution of peaks is more erratic, and peak strengths do not visibly seem to decay around a particular peak. Nevertheless, the Rising Sun Envelope Method can still be applied in this case.  

\section*{Discussion and Conclusions}
\noindent

The Rising Sun Envelope Method is accurate in locating peak intensity and peak location, allowing for discernment between \textit{true peaks} from \textit{random noise spikes}. The method takes into account the intrinsic noise of the measurement, which makes it suitable for  continuous, but noisy, functions, and in particular for experimental measurements. It  provides an automated approach to peak decomposition that uses the pointwise structure of the function and its intrinsic properties.

Our results also provide a more accurate way to locate the peak-edge region (see Figure \ref{fig:First_derivative}); overall, it is the first case in which this computation can be fully performed automatically.

The choice for peaks as interpolating points in polygonal or cubic spline approximations is shown to be an effective way to reduce the dimension of the spectral measurements. This choice is faithful to  two qualities that are crucial in material oxidation analysis:  peak location and its strength (see Figure \ref{fig:other_side_with_marked_peaks}). In effect, there are many reasons to pursue this kind of reductions: the first can be found in the  initial steps of the ML model in \cite{miyazato2019automatic}, where  a measure of the peak shift was the way the authors  found to  overcome this high dimensionality issue, considerably reducing the dimension of each XANES measurement in a way that is physically consistent with experiments; the Rising Sun Envelope method was originally developed with the intent of engineering more accurate horizontal peak shift features. Another reason to pursue dimensional reduction is due to   the difference between the dimension of spectral measurements (of order $\mathcal{O}(1000)$) and the
sample size (of order $\mathcal{O}(100)$), yielding  disparity  between number of variables and number of constraints  that prevents good fitting and the use of nonlocal Machine Learning methods (as Convolution Neural Networks) \cite[\S 9]{goodfellow2016deep}.
Still, a third reason can be presented: the availability of more data of XANES measurements by initiatives like  \cite{mathew2018high} will make the use ML and other Artificial Intelligence methods more common, accurate, and effective in the field of Material Sciences \cite{butler2018machine,takahashi2016materials}; the clamped spline interpolations we found not only reduce the dimension of the data, but further capture the complexity of the XANES measurement in a regularized fashion that is accurate and representative.\newline
\indent
Numerical code for this paper has been written in Python and is  available on \cite{github}.

\section*{Acknowledgment}\label{sec:acknowledgment}
This work is funded by Japan Science and Technology Agency(JST) CREST Grant Number JPMJCR17P2, JSPS KAKENHI Grant-in-Aid for Young Scientists (B) Grant Number JP17K14803, and Materials research by Information Integration (MI$^2$I) Initiative project of the Support Program for Starting Up Innovation Hub from JST.
XAFS spectrum data is utilized from XAFS database available in Institute of Catalysis, Hokkaido University \cite{Hokkaido}, and XAFS Spectra Library in the Center for Advanced Radiation Sources(CARS), the University of Chicago \cite{Chicago}.
R. M. is grateful to the  MathAM-Oil group (AIST/AIMR, Tohoku Univ.) for several comments and suggestions offered during a seminar where this project was presented, specially to A. Watanabe's (AIST) helpful comments on Raman spectra.




\textbf{Competing financial interests}\\
The authors declare no competing financial interests.

\appendix 
%
%
\begin{algorithm}[h]
\SetAlgoLined
\KwResult{Rising Sun Envelope Method for peak location of a spectral measurement}
\tcc{Forward passage}
 \KwData{Given a measurement $\mu(E)$ in the range $e_{-\infty}\leq  E \leq e_{+\infty}$, threshold parameters $(h_*^{(0)},d_*^{(0)})$, integers $M_{\text{after}}>0$ and  $M_{\text{before}}\geq 0$, and two sets of hyperparameters $\mathcal{H}_{\text{after}}$ and $\mathcal{H}_{\text{before}}$}
\While{$0\leq n <M_{\text{after}}$}{
  \eIf{ n is even}{
  Set $f(E)= \mu\Big|_{[e_n, e_{+\infty}]}(E)$ and compute the Rising Sund function $ \mathcal{R}_{f}(E)$  on energy range $e_n\leq E\leq e_{+\infty}$ \;
  \eIf{the first local maximum $\overline{E}$ of $\mathcal{R}_{f}(E)$ with threshold $(h_*^{(n)}, d_*^{(n)})$  exists}{Set $e_{n+ 1} \leftarrow \overline{E}$\;
  $n \leftarrow n+ 1$ and continue\;}{``only i-breakpoints could be found'', hence \textbf{break}\;}
  }{
  With  $f(E)$ as before, compute the Valley of Shadows function 
 $\mathcal{V}_{f}(E) = \mathcal{R}_{f}(E)- f(E)$; set $f(E)= \mathcal{V}_{f}\Big|_{[e_n, e_{+\infty}]}(E)$ \;
  \eIf{the first local maximum $\underline{E}$ of $f(E)$ with threshold $(h_*^{(n)}, d_*^{(n)})$  exists}{Set $e_{n+ 1} \leftarrow \underline{E}$\;
  $n \leftarrow n+ 1$ and continue\;}{``only i-breakpoints could be found'', hence \textbf{break}\;}
  }
   
 }
  \tcc{Reverse parrage: search on the left side of the 0th peak}
 \uIf{$M_{\text{before}}>0$}{
 Define $\widetilde{\mu}(E) = \mu\left(e_{-\infty} + e_{0} - E\right)$ in the range $e_{-\infty}\leq  E \leq e_{0} + e_{-\infty}$\;
 Run the previous code on threshold parameters $(\widetilde{h_*^{(0)}},\widetilde{d_*^{(0)}})$, $\widetilde{M_{\text{after}}}= M_{\text{before}}$,   $\widetilde{M_{\text{before}}}= 0$, \\
 and two sets of hyperparameters $\widetilde{\mathcal{H}_{\text{after}}}\equiv \mathcal{H}_{\text{before}}$ and $\widetilde{\mathcal{H}_{\text{before}}}\equiv \emptyset$\;
 
 }\textbf{end}
 
\caption{Rising sun Algorithm for peak decomposition. \label{alg:_peak_location}}
\end{algorithm}

 \bibliographystyle{ieeetr}


%

\end{document}